\def\targetformat{arXiv}		

\ifx\targetformat\undefined
	\def\clsstyle{prl} 
	\newcommand{\sectionheading}[1]{\paragraph{#1.}}
	\newcommand{\appref}[1]{the Supplemental Materials}
\else
	\def\clsstyle{pra} 
	\newcommand{\sectionheading}[1]{\section{#1}}
	\newcommand{\appref}[1]{App.~\ref{#1}}
\fi

\documentclass[aps,\clsstyle,reprint,twocolumn,showpacs,superscriptaddress]{revtex4-1}

\usepackage{amsmath, amssymb, braket, dsfont, times}
\usepackage{graphicx}
\usepackage{subfigure}		
\usepackage{epstopdf}
\usepackage{kbordermatrix} 
\renewcommand{\kbldelim}{(} 
\renewcommand{\kbrdelim}{)}
\usepackage[breaklinks]{hyperref}

\newcommand{\II}{{I{\mkern-4.5mu}I}}

\usepackage[usenames,dvipsnames]{color}			
\definecolor{Zcolour}{rgb}{0.992, 0.588, 0.22}
\definecolor{dkgreen}{rgb}{0,0.5,0}
\definecolor{purple}{rgb}{0.5,0,0.5}

\begin{document}

\title{Time-evolving a matrix product state with long-ranged interactions}
\author{Michael P. Zaletel}
\affiliation{Department of Physics, University of California, Berkeley, California 94720, USA}
\author{Roger S.~K. Mong}
\affiliation{Department of Physics and Institute for Quantum Information and Matter, California Institute of Technology, Pasadena, California 91125, USA}
\author{Christoph Karrasch}
\affiliation{Department of Physics, University of California, Berkeley, California 94720, USA}
\affiliation{Materials Sciences Division, Lawrence Berkeley National Laboratory, Berkeley, CA 94720, USA}
\author{Joel E. Moore}
\affiliation{Department of Physics, University of California, Berkeley, California 94720, USA}
\affiliation{Materials Sciences Division, Lawrence Berkeley National Laboratory, Berkeley, CA 94720, USA}
\author{Frank Pollmann}
\affiliation{Max-Planck-Institut f\"ur Physik komplexer Systeme, 01187 Dresden, Germany}
\date{\today}                                           
\begin{abstract}
We introduce a numerical algorithm to simulate the time evolution of a matrix product state under a long-ranged Hamiltonian.
In the effectively one-dimensional representation of a system by matrix product states, long-ranged interactions are necessary to simulate not just many physical interactions but also higher-dimensional problems with short-ranged interactions.
Since our method overcomes the restriction to short-ranged Hamiltonians of most existing methods, it proves particularly useful for studying  the dynamics of both power-law interacting one-dimensional systems, such as Coulombic and dipolar systems, and quasi two-dimensional systems, such as strips or cylinders.
First, we benchmark the method by verifying a long-standing theoretical prediction for the dynamical correlation functions of the Haldane-Shastry model.
Second, we simulate the time evolution of an expanding cloud of particles in the two-dimensional Bose-Hubbard model, a subject of several recent experiments.
\end{abstract}
\maketitle
\sectionheading{Introduction}

The ability to study  dynamical properties in and out of equilibrium is essential for the understanding of the physics of strongly interacting systems.
Following the success of the density-matrix renormalization group (DMRG) for finding one-dimensional (1D) ground states \cite{White-1992},
a number of closely related techniques have been developed to explore the dynamical properties of short-ranged 1D systems \cite{Vidal2003, White2004, Daley2004, Schmitteckert2004, Jeckelmann08}. 
This exciting development has given access to experimentally relevant observables, such as dynamical correlation functions which can be compared with data from neutron scattering and ultracold atomic gasses,  and non-equilibrium dynamics, providing insight into long standing questions about thermalization \cite{Schollwock2011}.
Simultaneously, large-scale DMRG has begun to study ground-state properties of quasi-two dimensional (2D) quantum systems, such as strips and cylinders, allowing one to probe much larger systems than accessible to exact diagonalization \cite{Stoudenmire2012}.
For example, DMRG studies provide solid evidence for the existence of a spin-liquid ground state in the kagome $S=1/2$ antiferromagnet \cite{Yan2011, Depenbrock2012}.
The 2D-DMRG method proceeds by ordering the sites of the 2D lattice into a 1D chain with long-ranged interactions.
Hopefully truly 2D tensor network methods will eventually supplant this approach \cite{Jordan2008, Corboz2011}, but currently DMRG remains a standard tool due to its reliability. 

\begin{figure}[t]
	\includegraphics[width=0.9\columnwidth]{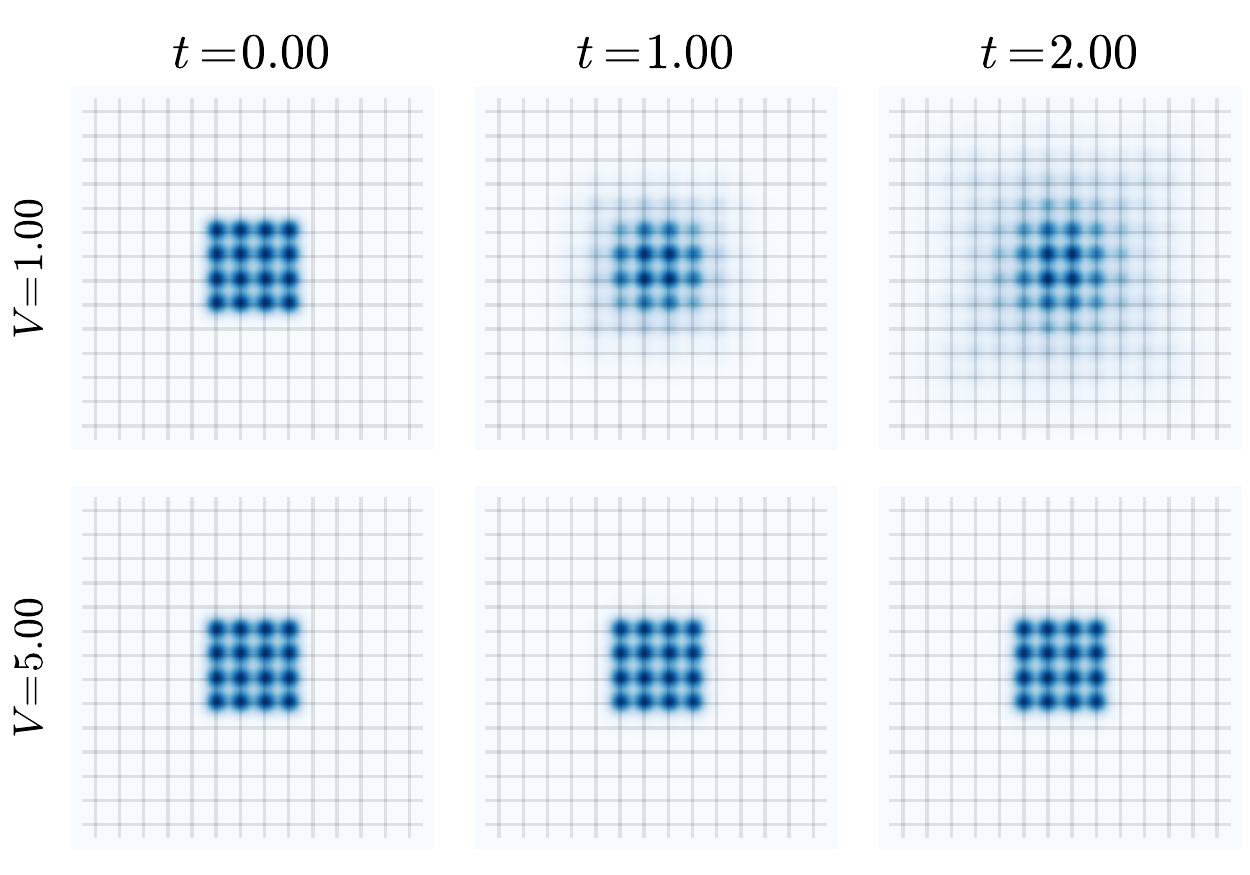}
	\caption{%
		Quasi-exact time evolution of interacting hard-core bosons in a $14\times14$ lattice trap.
		In addition to hopping of bandwidth $t = 1$, the bosons interact with nearest-neighbor repulsion $V$.
		16 bosons begin in an un-entangled product state, and evolve in time from left to right. In the top row, $V = 1$ and the bosons expand outward.
		In the bottom row, $t < V = 5$, the bosons remain trapped in a bound state due to the strong interactions.
		A similar effect has been observed experimentally in cold-atom optical lattices \cite{Winkler2006}.
	}
	\label{fig:2dbosons}
\end{figure}

It is now highly desirable to combine these two developments in order to evaluate dynamical properties of quasi-2D systems (e.g.,  the time evolution of bosons in a 2D optical trap as shown in Fig.~\ref{fig:2dbosons}).
However, the existing DMRG based time-evolution methods cannot be easily applied to a quasi-2D system.
This is mainly due to the long-ranged interactions that occur when representing a 2D system as a 1D chain; a similar difficulty exists for 1D systems with power-law Coulombic and dipolar interactions.

In this work we address this problem by providing a method to time-evolve long-ranged Hamiltonians.
The unique advantage of the method is that it simultaneously
	(a) can be applied to any long-ranged Hamiltonian while preserving all symmetries,
	(b) has a constant error per site in the thermodynamic limit at fixed computational effort,
	(c) can be applied to an infinitely long system assuming translation invariance
	and (d) can be easily implemented using standard DMRG methods.

Like other 1D methods, we work in the framework of matrix product states (MPSs) \cite{Fannes-1992,OstlundRommer1995,RommerOstlund1997}---%
	a variational ansatz for finitely-entangled states---within which we wish to simulate the full many-body dynamics (consequently, the method is practical only for moderately entangled systems). 
The structure of an MPS can be generalized to operators, called matrix product operators (MPO) \cite{Verstraete2004}.
An MPO can be very efficiently applied to an MPS using  standard methods \cite{Pirvu2010, Stoudenmire2010, Schollwock2011}.
If a long-ranged Hamiltonian $H$ has a compact MPO approximation for $e^{t H}$,
	then the time evolution can be efficiently simulated by successively applying the MPO to the MPS.
The most naive time-stepper, an Euler step $1 + t H$, as well as its Runge-Kutta \cite{White2004} and Krylov \cite{Schmitteckert2004, GarciaRipoll2006} improvements, indeed have an efficient MPO representation.
But these global methods have an error per \emph{site} which diverges with the system size $L$---for example as $\mathcal{O}(L t^2)$ for the Euler step---which eventually renders them impractical.
For certain simple $H$, such as a nearest neighbor interactions or a sum of commuting terms\cite{Pirvu2010}, a compact MPO with \emph{finite} error per site exists, which is the basis behind the highly successful time evolving block decimation (TEBD) \cite{Vidal2003} and tDMRG \cite{Daley2004}.
However, these methods do not generalize well for long-ranged Hamiltonian, which is the focus of this work.

The basic insight of this work is that a Hamiltonian which is expressed as a sum of terms $H =\sum_x H_x$ admits a \emph{local} version of a Runge-Kutta step; for instance we could improve the Euler step by taking
\begin{align}
	1 + t \sum_x H_x \rightarrow \prod_x (1 + t H_x).
	\label{eq:localeuler}
\end{align}
The error is still at $\mathcal{O}(t^2)$, so it is formally a 1\textsuperscript{st}-order time stepper.
But any set of distant regions all receive the correct 1\textsuperscript{st}-order step in parallel.
Hence, in contrast to the naive Euler step, the total error scales as $L t^2$, rather than as $L^2 t^2$. 
The main result of this work is that an improved version of Eq.~\eqref{eq:localeuler} has a very compact MPO representation which can easily be extended to higher-order approximations in $\mathcal{O}(t^p)$.

\begin{figure}[t]
	\includegraphics[width=0.975\columnwidth]{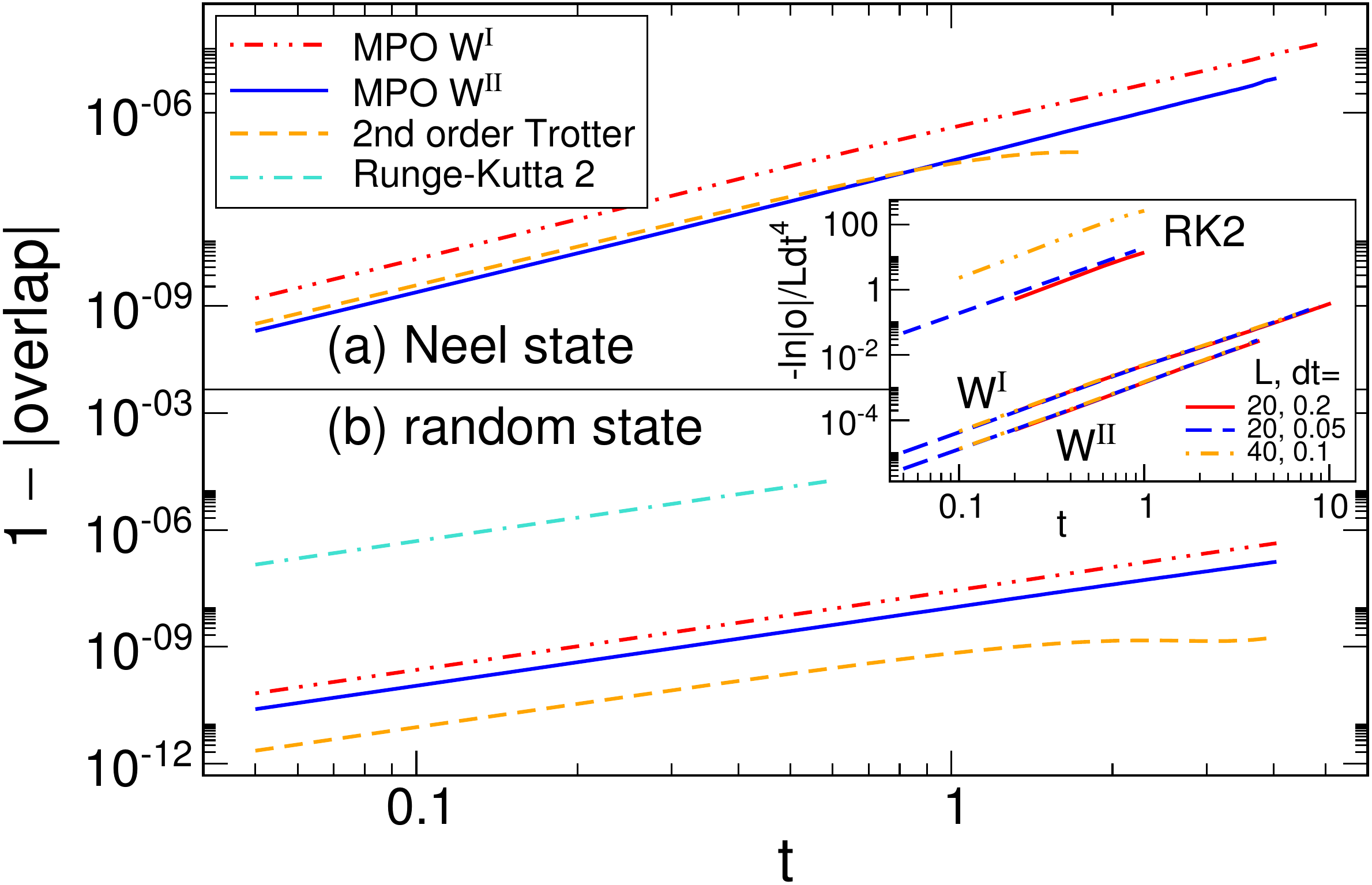}
	\caption{%
		Comparison of 2\textsuperscript{nd}-order MPOs $W^I$, $W^\II$, TEBD, and global Runge-Kutta for the spin-1/2 Heisenberg chain.
		4\textsuperscript{th}-order TEBD serves as a quasi-exact reference for calculating errors.
		Panels (a), (b) show quenches starting from a $L = 20$ Neel state and random state respectively.
		In the inset, we show the scaling of the errors for system sizes $L=20, 40$.
		For $W^{I/\II}$ we find perfect collapse to the expected scaling $L t^4$, as the error per site  remains constant in the thermodynamic limit.
		In contrast, for global Runge-Kutta the error increases as $L^5 t^4$.
	}
	\label{fig:comparison}
\end{figure}

In Fig.~\ref{fig:comparison}, we compare the accuracy of the methods proposed here, dubbed $W^{I}$ and $W^\II$, against TEBD and global 2\textsuperscript{nd} Runge-Kutta.
TEBD works for short-ranged Hamiltonians, so we compare by quenching from product states into the spin-1/2 nearest-neighbor Heisenberg chain, where a very high order TEBD calculation can serve as a quasi-exact reference. 
Runge-Kutta is orders of magnitude less accurate, with an error that scales as $L^5$ compared to $L$ for TEBD and $W^{I/\II}$. 
Both TEBD and $W^{I/\II}$ are comparable in accuracy; for evolution starting from a Neel state, $W^\II$ is slightly more accurate than TEBD, while from a random state TEBD is more accurate.
Any such difference can be easily mitigated by a small decrease in time step.
But unlike TEBD, $W^{I/\II}$ can be immediately applied to a long-ranged problem without a Trotter decomposition. 


To our knowledge, the other existing method which can time-evolve long-ranged interactions with a constant error per site is the recently developed time dependent variation principle (TDVP), which projects the exact Schr\"odinger equation into the MPS variational space and numerically integrates the resulting equations \cite{Haegeman2011, Dorando2009}.
While the method has yet to be applied to quasi-2D systems, a version was successfully applied to the long-ranged transverse field Ising model \cite{Hauke2013}.
However, in contrast to the proposal here, which involves the entirely standard tensor network technique of applying an MPO, the TDVP requires one to implement an entirely distinct and relatively complex set of algorithms.
It will be a useful subject of future work to make a detailed comparison between TDVP and this work.


The first application presented here is a calculation of a dynamical correlation function of the Haldane-Shastry spin chain, which is a 1D spin-half antiferromagnet with power law long-ranged interactions~\cite{Haldane88,Shastry88}.
Our numerical simulations not only agree with the analytic exact results \cite{HaldaneZirnbauer1993} up to long times, but also show a ballistic spreading of correlations consistent with the model's integrability; this also serves as a check of the method's accuracy.
The second application is the simulation of dynamics in a 2D Bose-Hubbard model. 
Here we focus on a class of experiments with ultracold atomic gases that study expansion of a cloud that is initially confined to a small region of the lattice \cite{Winkler2006}.
The main qualitative surprise in the experiments is that even repulsive interactions can lead to self-trapped states, which is reproduced in our model calculation along with several other features, shown in Fig~\ref{fig:2dbosons}.
We will further elaborate on these applications later.

\sectionheading{Matrix product operators}
In order to understand our main result, we review some basic facts regarding MPOs.
An operator $Z$ acting on a 1D chain with physical sites labeled by $i$ has an MPO representation  
\begin{align}
	Z &=  \cdots \hat{W}_{(1)} \hat{W}_{(2)} \hat{W}_{(3)} \cdots  \label{eq:suppress}
\end{align}
where each $\hat{W}_{(i)}$ is a matrix of operators acting on the Hilbert space of site-$i$
(with physical indicides $m_i,m_i'$),
\begin{align}
	[\hat{W}_{(i)}]_{a_{i-1} a_i} = \sum_{m_i,m_i'} [W_{(i)}]^{m_im_i'}_{a_{i-1} a_i} \ket{m_i} \bra{m_i'} ,
\end{align}
with $[W_{(i)}]^{m_im_i'}_{a_{i-1} a_i} \in \mathbb{C}$. In Eq.~\eqref{eq:suppress}, the matrices are contracted by summing over all indices $a_i=1,\dots,\chi_i$.
These indices live in the space between sites $(i,i+1)$, which refer to as a \emph{bond}.
The $\chi_i$s are called the MPO bond dimensions, and they denote the size of the $\hat{W}$ matrices.
Several algorithms have been developed for efficiently applying an MPO to an MPS, with effort of either $\mathcal{O}(\chi^2)$ or $\mathcal{O}(\chi^3)$ \cite{Pirvu2010, Stoudenmire2010, Schollwock2011}.

	Two classes are of interest to us; sums of local operators (such as a Hamiltonian), and exponentials of such sums (evolution operators).
We first review the structure of the former.
For the bond between sites $(i,i+1)$ that divides the system into regions $L_i$ and $R_i$, any Hamiltonian $H$ can be decomposed as
\begin{align}
	H &= H_{L_i} \otimes \mathds{1}_{R_i} +\mathds{1}_{L_i}\otimes H_{R_i} + \sum_{a_i=1}^{N_i} h_{L_i, a_i} \otimes h_{R_i, a_i} .
\label{eq:splitH}
\end{align}
Here $H_{L_i/R_i}$ are the components of the Hamiltonian localized purely to the left/right of the bond, while the $h_{L_i, a_i} \otimes h_{R_i, a_i}$ run over $N_i$ interaction terms which cross the bond.
There is a recursion between the decompositions on bond $(i-1,i)$ and $(i,i+1)$, which differ by the addition of site $i$.
\begin{align}
	\begin{pmatrix} H_{R_{i-1}} \\h_{R_{i-1}, a_{i-1}} \\ \mathds{1}_{R_{i-1}} \end{pmatrix}
	&= \kbordermatrix{ & 1 & N_{i} & 1 \\
			1 & \hat{\mathds{1}} & \hat{C} &\hat{D}    \\ 
			\!N_{i-1}\! & 0 & \hat{A}  &  \hat{B} \\ 
			1 & 0 & 0 & \hat{\mathds{1}} }_{\!\!(i)}
		\otimes \begin{pmatrix} H_{R_{i}} \\ h_{R_{i}, a_{i}} \\ \mathds{1}_{R_{i}} \end{pmatrix}
	\label{eqn:mpo}
\end{align}
Here $(\hat{A}, \hat{B}, \hat{C}, \hat{D})_{(i)}$ are matrices of operators acting on site $i$, with dimensions indicated on the border.
This recursion is in fact the MPO: the block matrix in the middle is $\hat{W}_{(i)}$, with size $\chi_i = N_i + 2$.
(See \appref{app:MPOexamples} for explicit examples of MPOs.)
The optimal $(\hat{A}, \hat{B}, \hat{C}, \hat{D})_{(i)}$ can be obtained using the block Hankel singular value decomposition, a well known technique in control theory known as balanced model reduction \cite{Kung1981}.
\begin{figure}[t]
	\includegraphics[width=0.95\columnwidth]{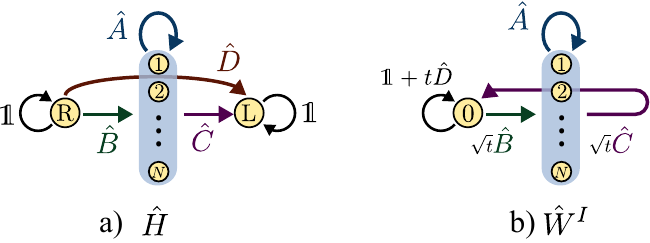}
	\caption{%
		Graphical depictions of MPOs for (a) the Hamiltonian $H$ and (b) the time-stepper $\hat{W}^I(t)$.
		As explained in Ref.~\onlinecite{CrosswhiteBacon2008}, by analogy to a finite-state-machine the indices of the MPO (labeling rows and columns) are represented as nodes of a graph,  while the entries of the MPO are edges.
}
	\label{fig:MPO} 
\end{figure}

We can view the recursion relation of Eq.~\eqref{eqn:mpo} as a finite state machine \cite{CrosswhiteBacon2008}; the transitions of the machine sequentially place the operators at each site, as illustrated in Fig.~\ref{fig:MPO}a.
The first/last indices of the MPO, which we denote by $\mathrm{L}$/$\mathrm{R}$ respectively, play a special role, as they indicate that no non-trivial operators have been placed to the left/right of the bond.
Due to the block-triangular structure of $\hat{W}$, once the MPO state transitions into the first index $\mathrm{L}$, it remains there in perpetuity, placing only the identity operator $\hat{\mathds{1}}$  with each $\hat{W}$.
The transition from $\mathrm{R}$ to $\mathrm{L}$ (not necessarily in one step) places some local operator $H_x$; the sum over all such paths generates the Hamiltonian.

\sectionheading{Time evolution operators}
Given the sum of terms $H =\sum_x H_x$, our goal is to find an efficient MPO approximation for
\begin{align}
	U(t) = 1 + t \sum_x H_x + \frac{1}{2} t^2 \sum_{x, y} H_x H_y + \cdots.
	\label{eq:U}
\end{align}
In the most general case, an approximation for $U(t)$ is necessary, which brings us to our main result.

While the local Euler  step defined in Eq.~\eqref{eq:localeuler} does not have a simple MPO approximation, a slight modification does.
Let us define $x < y$ if the sites affected by $H_x$ are strictly to the left of those affected by $H_y$.
Consider an evolution operator which keeps all non-overlapping terms:
\begin{align}
	U^{I}(t) &= 1 + t \sum_x H_x + t^2 \sum_{x < y} H_x H_y
	\label{eq:U1}
		\\ &\quad
		+ t^3 \sum_{x < y < z} H_x H_y H_z + \dots
	\notag
\end{align}
These contributions are a subset of Eqs.~\eqref{eq:localeuler} and \eqref{eq:U}.
The first error occurs at order $t^2$, for terms $H_x, H_y$ which overlap on at least one site.
For a system of length $L$, there are $\mathcal{O}(L)$ such terms, so the error is $\mathcal{O}( L t^2 )$.
Hence the error is constant per \emph{site}.
Remarkably, $U^I$ has an exact compact MPO description ``$W^I$'', and is trivial to construct from the $(A, B, C, D)$ of $H$, illustrated in Fig.~\ref{fig:MPO}b.
It has a block structure of total dimension $\chi_i = N_i + 1$:
\begin{align}
	\hat{W}^I_{(i)}(t) = \kbordermatrix {& 1 &  N_{i} \\
		1 & \hat{\mathds{1}}_{(i)} + t \hat{D}_{(i)}  & \sqrt{t} \hat{C}_{(i)}   \\ 
		\!N_{i-1}\! &   \sqrt{t} \hat{B}_{(i)}  &  \hat{A}_{(i)} }.
	\label{eq:W1}
\end{align}

While $\hat{W}^I$ is trivial to construct and performs well, it is not quite optimal.
For example, a Hamiltonian consisting of purely onsite terms has a trivial MPO representation for $e^{tH}$, since the evolution is just a tensor product.
Yet the MPO constructed from $\hat{W}^I$ would only produce the approximation $U^I = \prod_x (1 + t H_x)$  in this case.

We propose an improvement to Eq.~\eqref{eq:U1}, where we also keep terms which may overlap by one site.
Let $\braket{x,\dots,z}$ denote a collection of terms in which no two cross the same bond.
Arbitrarily high powers of a single site term, for example, can appear in these collections.
Consider an evolution operator which keeps all such terms:
\begin{align}
	U^\II(t) &= 1 + t \sum_x H_x + \frac{t^2}{2} \sum_{\;\braket{x, y}} H_x H_y 
		\\&\quad + \frac{t^3}{6} \sum_{\;\braket{x, y, z}} H_x H_y H_z + \dots\ 	.\notag
\end{align}
Again, the first error occurs at $t^2$, with $L$ such terms, so the error is still formally $\mathcal{O}( L t^2 )$.
But for typical interactions far fewer terms are dropped than in $U^I$; in particular since any onsite term does not self-overlap across any bond, they are captured to \emph{all} orders.
While there isn't an exact compact MPO representation for $U^\II$, we can construct an MPO approximation $\hat{W}^\II$ which differs $U^\II$ by $\mathcal{O}(L t^3)$.
Because the different is at higher order than the accuracy of $U^\II$, $\hat{W}^\II$ still gives a noticeably better approximation than $\hat{W}^I$, and retains the feature that an onsite interaction is kept exactly.

	The MPO $\hat{W}^\II$ is more complicated to construct, so for a detailed derivation of $\hat{W}^\II$ and an algorithm to compute it we refer to \appref{app:w2}. 
It takes the form
\begin{align}
	\hat{W}^\II = \kbordermatrix {& 1 &  N_{i} \\
		1 & \hat{W}_D^\II  & \hat{W}_C^\II  \\ 
		\!N_{i-1}\! &  \hat{W}_B^\II   &  \hat{W}_A^\II  }.
\end{align}
To define the sub-blocks, introduce two vectors of formal parameters $\phi_{a}, \bar{\phi}_b$, with $a = 1, \ldots, N_{i-1}$, $b = 1, \ldots, N_i$.
Let $\phi\cdot\hat{A}_{(i)}\cdot\bar{\phi}$ denote a dot product of these formal parameters into the MPO indices of $\hat{A}_{(i)}$.
The sub-blocks are defined by a Taylor expansion in terms of $\phi$, $\bar\phi$,
\begin{align}
	& e^{  \phi\cdot\hat{A}\cdot\bar{\phi}  + \phi\cdot\hat{B} \sqrt{t} +  \sqrt{t} \hat{C}\cdot\bar{\phi} + t \hat{D} } \\
	&\quad = \hat{W}_D^\II + \hat{W}_C^\II\cdot\bar{\phi} + \phi\cdot\hat{W}^\II_B  + \phi\cdot\hat{W}_A^\II\cdot\bar{\phi} + \dots \notag
\end{align}
Notice $\hat{W}_D^\II = e^{t \hat{D}}$ is simply the onsite term, which is kept exactly.
We also note that $H$ has many different MPO representations, and at 2\textsuperscript{nd}-order $\hat{W}^\II$ is not invariant under different choices.
This choice can be exploited to further reduce errors (cf.\ \appref{app:MFMPO}).
Finally, if $H$ is a sum of commuting (or anticommuting) terms, there is an analytic MPO representation for $e^{tH}$ given in \appref{app:mpocommute}.

	As with TEBD, we want to construct approximations with errors at higher order $\mathcal{O}(L t^p)$ in $t$, which allow one to use much larger time steps.
In fact, simply by cycling through a carefully chosen set of step constants $\{t_a\}$ we can obtain approximations of arbitrarily high order.
Each stage of the approximation should have a compact MPO expression (otherwise the increased complexity cancels the gains of a larger time step),  so we consider an ansatz of the form
\begin{align}
	U^I(t_1) U^I(t_2) \cdots U^I(t_n) = U(t) + \mathcal{O}(L t^p),
	\label{eq:order}
\end{align}
where $p-1$ is the approximation order.
Our goal is to determine a set of step constants $\{t_a\}$ which produce the desired order.
For example, to find a 2\textsuperscript{nd}-order step ($p = 3$), we expand Eq.~\eqref{eq:order} order by order and find constraints
\begin{align}
	\sum_a t_a  = t, \quad \sum_{a < b} t_a t_b = \frac{1}{2} t^2, \quad \sum_a t^2_a = 0
\end{align}
which can be solved by $t_1 = \frac{1+i}{2}t$, $t_2 = \frac{1-i}{2}t$.
One can continue to arbitrary order; a set of 4 $t_a$'s is required at 3\textsuperscript{rd} order, a set of 7 at 4\textsuperscript{th} order.
	Thus, by alternating between two compact MPOs, $W^I(t_1)$ and $W^I(t_2)$, we obtain a 2\textsuperscript{nd}-order approximation, and likewise for $W^\II$.
As shown in Fig.~\ref{fig:comparison}, the 2\textsuperscript{nd}-order  behavior is preserved even when truncation to the MPS ansatz intervenes between steps, so the 2\textsuperscript{nd}-order time step is no more demanding than the 1\textsuperscript{st}-order one.

\begin{figure}[tb]
	\includegraphics[width=0.95\columnwidth]{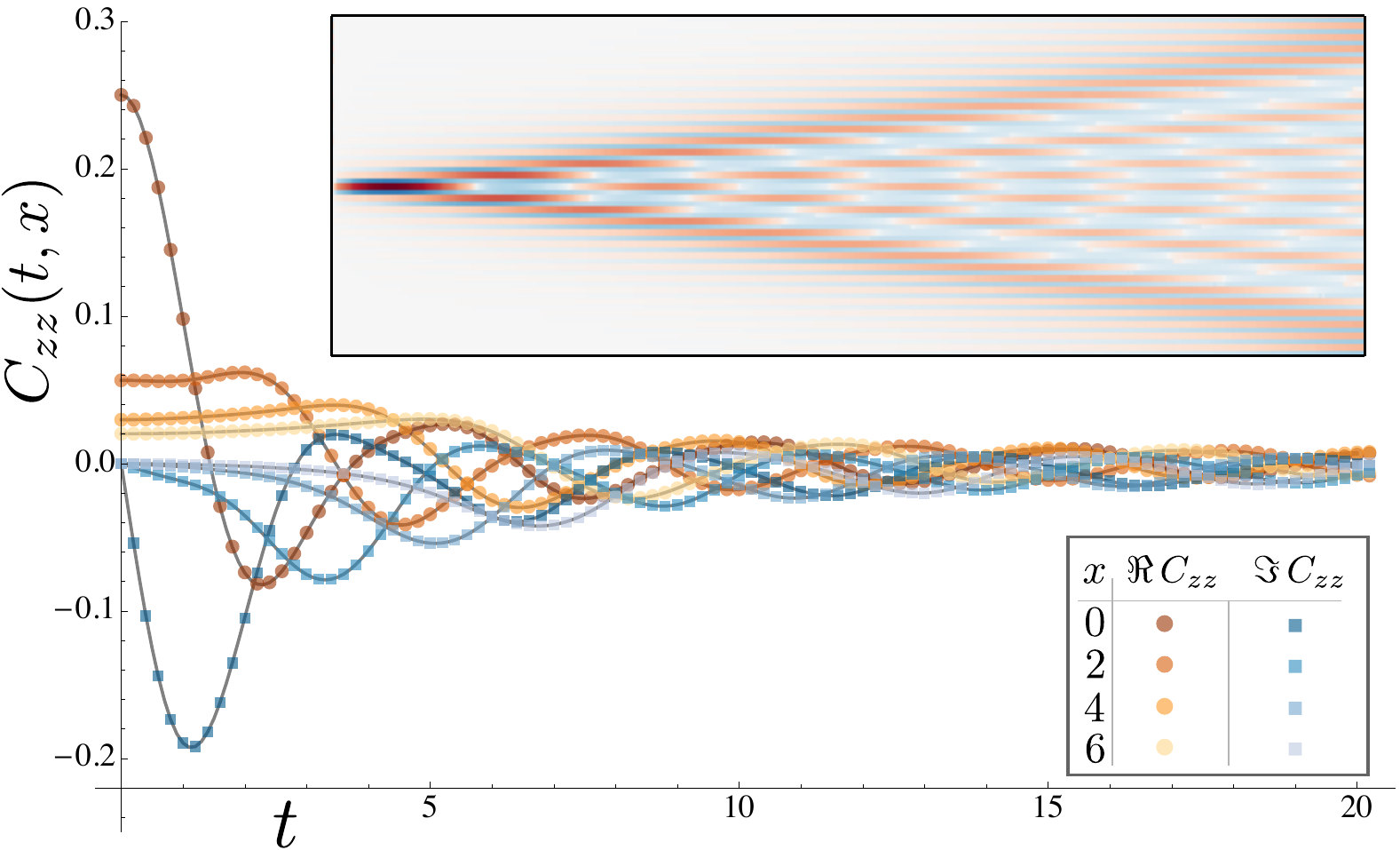}
	\caption{%
		Time evolution of the response function $C_{zz}(t, x) = \bra{0} S^z(t, x) S^z(0, 0) \ket{0}$ for the Haldane-Shastry model.
		Discrete data points are evaluated numerically using the 2\textsuperscript{nd}-order MPO time stepper $W^\II$ ($dt = 0.025$), shown here for positions $x = 0, 2, 4, 6$.
		The model is exactly solvable, with the analytic prediction shown in solid curves, giving beautiful agreement with the MPO.
		The inset shows a density plot of  $C_{zz}(t, x)$ in the $t\mbox{--}x$ plane.
	}
	\label{fig:czz}
\end{figure}

\sectionheading{Applications}
Our first system beyond the reach of TEBD is the spin-1/2 Haldane-Shastry model, an exactly solvable critical spin chain with long-ranged Hamiltonian 
\begin{align}
	H_{\textrm{HS}} = \sum_{x, r > 0} \frac{\mathbf{S}_x \cdot \mathbf{S}_{x+r}}{r^2}.
\end{align}
The model can be viewed as a lattice form of the Calogero-Sutherland continuum model of fractional statistics~\cite{calogero,sutherland} and is connected to the Laughlin fractional quantum Hall wavefunction with an exact MPS representation~\cite{sierracirac}.
The dynamical correlation function
\begin{align}
	C_{zz}(t, x) = \Braket{ S^z(t, x) S^z(0, 0) }
\end{align}
was first calculated analytically by Haldane and Zirnbauer~\cite{HaldaneZirnbauer1993}.
As the system is critical and the Hamiltonian long-ranged, numerically obtaining $C_{zz}$ is a stringent test of the proposed method.
We use an MPO approximation of the Hamiltonian to capture the $r^{-2}$ power law with high accuracy out to about 200 sites \cite{Crosswhite2008}.
After using infinite DMRG~\cite{McCulloch-2008, Crosswhite2008, Kjall-2013} to obtain the ground state with infinite boundary conditions, we act with $S^z$ and time evolve via $W^\II$.
As described in Fig.~\ref{fig:czz}, the numerically computed $C_{zz}$ is nearly identical to the analytic prediction (\appref{app:HS}) out to significant time scales.


Finally, one of the most interesting potential applications is time-evolving finitely-entangled 2D systems.
We make a preliminary study by considering the 2D Bose-Hubbard model with a hard-core interaction and nearest neighbor repulsion $V$.
Recently there have been several experimental and theoretical studies of the expansion of strongly-interacting clouds~\cite{Winkler2006, Ronzheimer2013,Ganahl-2013}.
A particularly counterintuitive result is that in a closed system with a periodic potential, \emph{repulsive} interactions can generate many-body bound states. 
This is because when the repulsion $V$ exceeds the bandwidth $t$, there is no way for the interaction energy  to transform into kinetic energy. The same effect occurs for strong attractive interactions. 
This effect is seen experimentally in anisotropic Bose-Hubbard models, where the repulsion is an onsite $U$ \cite{Winkler2006}.
Here we let a 16-boson $n=1$ product state expand into a $14\times14$ grid.
As shown in Fig.~\ref{fig:2dbosons}, the repulsion $V$ has a dramatic effect on the expansion, trapping the bosons into a bound state.
Because the 2D lattice has been turned into a 1D chain, the errors in $W^\II$ are \emph{highly} anisotropic.
Nevertheless we find that with a time step $dt = 0.01$, the density remains rotationally symmetric to within 4\% at $t = 2$.

\sectionheading{Conclusion}

To conclude, we have introduced a matrix-product operator based  algorithm to simulate the time-evolution of a matrix-product state under a long-ranged Hamiltonian. The method was first benchmarked against exact results: (i) We compared to results of existing numerical methods  for 1D short ranged models.  (ii) For the long-ranged Haldane-Shastry model, we verified the theoretical prediction for the dynamical correlation functions. We then presented results of a preliminary study of the expansion of interacting bosons in a 2D trap. Given the recent successes of DMRG  for investigating gapped 2D ground state and their gapless edges,  the techniques presented here could open the door to numerically calculating experimentally relevant dynamic quantities such as spectral functions.

\begin{acknowledgments}
We are grateful to D.~Varjas and J.~H.~Bardarson for helpful conversations.
The authors wish to thank
	NSF DMR-1206515 (M.Z. and J.E.M.),
	the Sherman Fairchild Foundation (R.M.),
	the Nanostructured Thermoelectrics program of DOE BES (C.K.),
	and the Simons Foundation (J.E.M.).
\end{acknowledgments}

\bibliography{time_evolution}

\ifx\targetformat\undefined
\else
	\clearpage
	\appendix
	\section{MPO examples}
\label{app:MPOexamples}

In this section, we provide explicit examples of MPOs for pedagogical purposes.

To reiterate from the main text, an MPO describes an operators written as a product of $\hat{W}$'s
\begin{align}
	\cdots \hat{W}_{(1)} \hat{W}_{(2)} \hat{W}_{(3)} \cdots  ,
\end{align}
where each $\hat{W}_{(i)}$ is a matrix of operators acting on site $i$.
An MPO for a Hamiltonain can always be casted in the form
\begin{align}
	\hat{W}_{(i)} &= \kbordermatrix{ & 1 & N_{i} & 1 \\
			1 & \hat{\mathds{1}} & \hat{C} &\hat{D}    \\ 
			\!N_{i-1}\! & 0 & \hat{A}  &  \hat{B} \\ 
			1 & 0 & 0 & \hat{\mathds{1}} }_{\!\!(i)} .
\end{align}
$\hat{D}$ is simply an operator, $\hat{C}$ and $\hat{B}$ are, respectively, a row and column vector, an $\hat{A}$ is an $N_{i-1} \times N_i$ matrix of operators.

Consider the transverse field Ising model with Hamiltonian
\begin{align}
	H_\textrm{TFI} = -J \sum_i \hat{Z}_i \hat{Z}_{i+1} - h \sum_i \hat{X}_i,
	\label{eq:H_tfi}
\end{align}
where $\hat{X}$ and $\hat{Z}$ are Pauli operators.
This Hamiltonian may be constructed as an MPO with
\begin{align}
	\hat{W}_{(i)} = \begin{pmatrix}
			\hat{\mathds{1}} & \hat{Z} & -h\hat{X} \\
			0	&	0	&	-J\hat{Z}	\\
			0	&	0	&	\hat{\mathds{1}}
		\end{pmatrix}_{\!\!(i)} .
\end{align}
Hence $N_i = 1$ for all bonds, and the MPO has bond dimension $\chi_i = 3$.
We can also read off the $(\hat{A},\hat{B},\hat{C},\hat{D})$ operators as $(0, -J\hat{Z}, \hat{Z}, -h\hat{X})$.
We note that this MPO is not unique for Hamiltonian Eq.~\eqref{eq:H_tfi} (cf.\ App.~\ref{app:MFMPO}).
Due to the absence of $\hat{A}$, the Hamiltonian consists of only onsite and nearest-neighbor terms.
Here $\hat{D}$ always denote the onsite term, and the pair terms are given by $\hat{C}_i \hat{B}_{i+1}$.

Our second example is a long-ranged XY-chain, with exponentially decaying couplings.
\begin{align}
	H = J \sum_{i<j} e^{-\alpha|i-j|} \big( \hat{X}_i \hat{X}_j + \hat{Y}_i \hat{Y}_j \big).
\end{align}
A corresponding MPO with $N_i = 2$ is as follows,
\begin{align}
	\hat{W}_{(i)} = \begin{pmatrix}
			\hat{\mathds{1}} & e^{-\alpha}\hat{X} & e^{-\alpha}\hat{Y} & 0 \\
			0 & e^{-\alpha}\hat{\mathds{1}} & 0 & J\hat{X} \\
			0 & 0 & e^{-\alpha}\hat{\mathds{1}} & J\hat{Y} \\
			0 & 0 & 0 & \hat{\mathds{1}}
		\end{pmatrix}_{\!\!(i)} .
\end{align}
Here $\hat{A}$ is a non-trivial $2\times2$ matrix of operators, which allows terms to reach beyond two neighboring sites.
Each insertion of the $\hat{A}$ matrix increases the separation of the bookends $\hat{X}/\hat{Y}$ by 1 site, and also reduces its amplitude by $e^{-\alpha}$ factor.

\section{\texorpdfstring{Computing $W_\II$}{Computing}}
\label{app:w2}
We defer the derivation of $\hat{W}^\II$ until after App.~\ref{app:mpocommute}, but first give an algorithm to compute it.
We must compute objects of the form
\begin{align}
\label{eq:appw2}
\hat{W}[\phi, \bar{\phi}] &= e^{  \phi\cdot\hat{A}\cdot\bar{\phi}  + \phi\cdot\hat{B} \sqrt{t} +  \sqrt{t} \hat{C}\cdot\bar{\phi} + t \hat{D}} \\
 &= \hat{W}_D + \hat{W}_C\cdot\bar{\phi} + \phi\cdot\hat{W}_B  + \phi\cdot\hat{W}_A\cdot\bar{\phi} + \cdots
\notag
\end{align}
For certain cases where the Hamiltonian is free, so that $A$ contains no field operators, $B, C$ are linear in field operators, and $D$ is quadratic in field operators, the result can be obtained using Pfaffians or permanents for fermionic and bosonic theories respectively. Here we discuss only the most general case, where the result must be obtained numerically.

	Let's compute $\hat{W}_{A; a \bar{a}}$, where  $a, \bar{a}$ index the rows and columns in correspondence with $\phi_a, \bar{\phi}_{\bar{a}}$.
At this order, we can consider $\phi_a, \bar{\phi}_{\bar{a}}$ to be formal objects defined by the property $\phi_a^2 = \bar{\phi}_{\bar{a}}^2 = 0$, and they commute with all other objects.
For computational purposes, we can then represent  $\phi_a$ as a hard-core boson creation operator $\phi_a \to c^\dagger_a$, and likewise $\bar{\phi}_{\bar{a}} \to \bar{c}^\dagger_{\bar{a}}$, restricted to an occupation of at most 1 $c$-type and 1 $\bar{c}$-type boson.
We denote the Hilbert space of the $c / \bar{c}$ type bosons by $\mathcal{H}_{c / \bar{c}}$, and $\mathcal{H}_\textrm{phys}$ the Hilbert space of the physical site.
The desired entries of $\hat{W}_A$, which are \emph{operators} in $\mathcal{H}_\textrm{phys}$, can be obtained by calculating a vacuum expectation values in the Hilbert space of the $\mathcal{H}_{c / \bar{c}}$ coupled to the physical site:
\begin{align}
\hat{W}_{A; a \bar{a}} &= \bra{0, \bar{0}} c_a \bar{c}_{\bar{a}} e^{  c^\dagger\cdot\hat{A}\cdot\bar{c}^\dagger + c^\dagger\cdot\hat{B} \sqrt{t} +  \sqrt{t} \hat{C}\cdot\bar{c}^\dagger + t \hat{D}} \ket{0, \bar{0}} \\
&= \bra{0, \bar{0}} c_a \bar{c}_{\bar{a}}  e^{  c^\dagger_a \bar{c}^\dagger_{\bar{a}}\hat{A}_{ab} + c^\dagger_a \hat{B}_a \sqrt{t} +  \sqrt{t} \hat{C}_b \bar{c}^\dagger_{\bar{a}} + t \hat{D}} \ket{0, \bar{0}}
	\notag
\end{align}	
To be more explicit, the argument of the exponential is an operator in the space $\mathcal{H}_c \otimes \mathcal{H}_{\bar{c}} \otimes \mathcal{H}_\textrm{phys}$.
The desired entry $\hat{W}_{A; a, \bar{a}}$ is the transition amplitude from the vacuum $\ket{0, \bar{0}}$ of the $\mathcal{H}_c \otimes \mathcal{H}_{\bar{c}}$ into the occupied state $\bra{0, \bar{0}} c_a \bar{c}_{\bar{a}}$.
Because the operators are restricted to single occupation, $c^2 = \bar{c}^2 = 0$, when computing the particular entry $\hat{W}_{A; a \bar{a}}$  we only need the Hilbert space of two hard-core bosons $c_a, \bar{c}_{\bar{a}}$ as well as the physical Hilbert space of a single site; if the latter dimension is $d$, the total dimension is $2^2 d$.
Thus the matrix elements can be obtained by exponentiating a matrix of dimension $4 d$, which is trivial.
This is repeated for the $N^2$ entries of $\hat{W}_{A; a \bar{a}}$.
Results for $\hat{W}_{B; a}$ follow as a byproduct by calculating the transition into  $\bra{0, \bar{0}} c_a$, and similarly for $C, D$.

	All together, $\hat{W}^\II$ can be computed with complexity $\mathcal{O}(N^2 d^3)$.

\section{Exact MPO exponentiation for commuting Hamiltonians}
\label{app:mpocommute}
	Here we obtain the exact MPO description for $e^{H}$ when $H$ is a sum of commuting terms such as $\sum_{i, j} \hat{X}_i \hat{X}_j t_{ij}$.
This result generalizes the nearest-neighbor case investigated in Ref.~\onlinecite{Pirvu2010}.
Specifically, we address the case in which $\hat{A}, \hat{B}, \hat{C}, \hat{D}$ must all commute.

Suppose the data $(A, B, C, D)_{(i)}$ of the MPO representation for $H$ is given, with bond dimensions $\chi_i = 2 + N_i$.
On each bond $(i,i+1)$, introduce a vector of complex fields $\phi_i = (\phi_{i, 1}, \dots, \phi_{i, N_i})$, with complex conjugate $\bar{\phi}_i$ and indices $a_i = 1, \dots, N_i$ in correspondence with the non-trivial MPO indices in $H$.
(That is, any MPO indices that is not $\mathrm{L}$ or $\mathrm{R}$.)
Using the fundamental rule of complex Gaussian integrals, 
\begin{align}
	\frac{1}{\pi} \int\!d^2\phi\,  e^{-\bar{\phi}\phi + J\bar{\phi} + \phi\bar{J}} = e^{ J\bar{J} },
\end{align}
the exponential factors as
\begin{align}
	e^{H} &= \int\! \mathcal{D}[\phi_i,\bar\phi_i] \; e^{ H_{L_i} + h_{L_i}\cdot\bar{\phi}_i } e^{-\bar{\phi}_i\cdot\phi_i} e^{\phi_i\cdot h_{R_i} + H_{R_i} }
\end{align}
where the dot-product is the sum $\sum_{a_{i} = 1}^{N_i}$, and $\mathcal{D}[\phi_i,\bar\phi_i]$ is shorthand for $\prod_{a_i} (d^2\phi_{i,a_i} / \pi)$.
This identity requires that all terms commute; otherwise discrepancies arise at second-order in $H$.

\begin{widetext}
Now using the MPO recursion of Eq.~\eqref{eqn:mpo}, we can peal off one site:
\begin{align}
	H_{R_i} + \phi_i\cdot h_{R_i} = \phi_i\cdot \hat{A}_{i+1}\cdot h_{R_{i+1}} + \phi_i\cdot \hat{B}_{i+1} +  \hat{C}_{i+1}\cdot h_{R_{i+1}} + \hat{D}_{i+1} + H_{R_{i+1}}.
\end{align}
Thus if we introduce a new vector of fields $\phi_{i+1, a_{i+1}}$ which runs over $a_{i+1} = 1, \dots, N_{i+1}$, we can write
\begin{align}\begin{split}
	e^{\phi_i\cdot h_{R_i} + H_{R_i} }  &= \int\! \mathcal{D}[\phi_{i+1},\bar\phi_{i+1}] \;
		\hat{U}_{\phi_i, \bar{\phi}_{i+1}}  e^{-\bar{\phi}_{i+1} \phi_{i+1}}e^{ \phi_{i+1}\cdot h_{R_{i+1}} + H_{R_{i+1}} } ,		\\
	&\textrm{where}\quad
	\hat{U}_{\phi_i, \bar{\phi}_{i+1}} \equiv e^{  \phi_i\cdot \hat{A}_{i+1}\cdot \bar{\phi}_{i+1}  + \phi_i\cdot \hat{B}_{i+1} +  \hat{C}_{i+1}\cdot \bar{\phi}_{i+1} + \hat{D}_{i+1} } .
	\label{eq:bSVD}
\end{split}\end{align}
By repeating this step on all the bonds, we find
\begin{align}
e^{H} = \int \mathcal{D}[\phi, \bar{\phi}] \left[ \cdots  e^{-\bar{\phi}_{i} \phi_{i}} \hat{U}_{\phi_i, \bar{\phi}_{i+1}}  e^{-\bar{\phi}_{i+1} \phi_{i+1}} \hat{U}_{\phi_{i+1}, \bar{\phi}_{i+2}} \cdots \right] .
	\label{eq:coh_mpo}
\end{align}
\end{widetext}
This is a matrix product operator in which the auxiliary bonds are labeled by a set of continuous numbers $\phi_i$, rather than discrete indices; it is a ``coherent state MPO.''
To bring the result to a discrete form, we note that an integral of the form Eq.~\eqref{eq:coh_mpo} is a discretized coherent state path integral for $N_i$ bosons, so the integrals can be converted to discrete sums over the \emph{many-body} Hilbert space of $N_i$ bosons. The basic manipulation is the Taylor expansion:
\begin{align}
	Y_{\phi} \equiv \sum_{n=0}^{\infty} Y_n \frac{\phi^n}{\sqrt{n!}} &\quad\textrm{(and likewise for any  tensor)} \\
\notag
	\frac{1}{\pi} \int\! d^2\phi\, X_{\bar{\phi}} e^{-\bar{\phi} \phi } Y_{\phi}
		&= \frac{1}{\pi} \sum_{\bar{n}, n} X_{\bar{n}} Y_n \, \int\! d^2\phi\, \frac{\bar{\phi}^{\bar{n}} \phi^{n}}{\sqrt{\bar{n}! \, n!}} e^{-\bar{\phi}\phi } \\
		&= \sum_n X_{n} Y_n
\end{align}
The integer $n$ is the `occupation.' Note that if a tensor depends on multiple variables (such as the vector $\phi_{i, a_i}$ ), then the above rule extends via a simple product. So if we define a vector of occupations $n_i = (n_{i, 1}, \dots, n_{i, N_i})$, whose values index the Hilbert space of $N_i$ bosons, we can Taylor expand $U$ as
\begin{align}
\hat{U}_{\phi_i, \bar{\phi}_{i+1}} \equiv  \sum_{\{n_i\}, \{ \bar{n}_i \}} \hat{U}_{ n_i, \bar{n}_{i+1} } \frac{ \phi_i^{n_i} \bar{\phi}_{i+1}^{\bar{n}_{i+1}}  }{\sqrt{ |n_i!||\bar{n}_{i+1}!|}}
\end{align}
with $|n_i!| = \prod_{a_i} (n_{i, a_i})!$.
The MPO for the exponential is
\begin{align}
e^{H} = \sum_{ \{n_i\}} \left[ \cdots \hat{U}_{ n_i, n_{i+1} } \hat{U}_{ n_{i+1}, n_{i+2} } \cdots \right]
\end{align}
Now in principle each sum on the bonds is over the many-body Hilbert space of $N_i$ bosons, which is infinite. But there will be ``Boltzmann factors'' associated to these states which allows for a sensible truncation.

Furthermore, in certain situations, such as for a nearest-neighbor interaction of Pauli-matrices, $H = \sum_i \hat{X}_i \hat{X}_{i+1}$, $\hat{U}_{ n_{i+1}, n_{i+2} }$ only has rank 2, resulting in the $\chi = 2$ MPO reported previously \cite{Pirvu2010}.

\section{\texorpdfstring{Derivation of $\hat{W}^\II$}{Derivation}}
\label{app:derw2}
Comparing Eq.~\eqref{eq:appw2} with Eq.~\eqref{eq:bSVD}, we see that $\hat{W}^\II$ is precisely a truncation of $\hat{U}_{n_i, n_{i+1}}$ to an occupation of at most a single boson on each bond.
The occupation number of bosons across a bond encodes the number of terms in the Hamiltonian which cross the bond in the Taylor expansion of $e^{\sum_x H_x}$.
Hence by truncating $\hat{U}$ to a  maximum occupation of 1, we keep all non bond-overlapping terms. 
However, in the derivation of the exact MPO $e^{tH}$ we required all terms to commute. 
Careful inspection shows that the non-commutivity only shows up at 3\textsuperscript{rd}-order in $H$.
Hence in general $\hat{W}^\II$ is only an approximation to the sum of all non bond-overlapping terms, with errors at $\mathcal{O}(t^3)$.
But these errors are subleading in comparison to the terms dropped (by the truncation) at $\mathcal{O}(t^2)$, so are unimportant.

\section{Taking advantage of different MPO decompositions}
\label{app:MFMPO}
There are numerous ways to decompose a Hamiltonian as $H = \sum_x H_x$, and hence many decompositions into an MPO.
For instance, a ferromagnetic interaction can be written as 
\begin{align}
	H_F &= -\sum_i \hat{Z}_i \hat{Z}_{i+1} \\ &= - \sum_i \left[ (\hat{Z}_i - h) (\hat{Z}_{i+1} - h) + 2 h \hat{Z}_i - h^2 \right] \notag
\end{align}
with MPO
\begin{align}
	\hat{W}^{H_F} = \begin{pmatrix}1 & -\hat{Z} & 0 \\0 & 0 & \hat{Z} \\0 & 0 & 1\end{pmatrix} \textrm{ or } \begin{pmatrix}1 & -(\hat{Z}-h) & h^2 - 2 h \hat{Z} \\0 & 0 & (\hat{Z}-h) \\0 & 0 & 1\end{pmatrix}.
\label{eq:mpo_shift}
\end{align}
The MPO $\hat{W}^\II$ is \emph{not} invariant under such shifts (at 2\textsuperscript{nd}-order). 
This can be used to improve the effective accuracy of $\hat{W}^\II$.

In principle one could try to optimize over all the MPO representations of $H$ in order to minimize the error in $\hat{W}^\II$.
It is an open question whether there is a practical method to do this.
As a toy model we compute the error $| (U(dt) - W^\II(dt)) \ket{\psi} |$ for the ferromagnet $H_F$ as a function of the shift $h$ given in Eq.~\eqref{eq:mpo_shift}.
To leading order,
\begin{align}\begin{split}
&	\Big| (U(dt) - W^\II(dt)) \ket{\psi} \Big|^2
\\&\quad		\propto dt^2 \sum_i \bra{\psi} (\hat{Z}_i - h)^2 (\hat{Z}_{i+1} - h)^2 \ket{\psi}, 
\end{split}\end{align}
since $W^\II$ drops these two-site terms  at 2\textsuperscript{nd}-order.
So, in principle, the optimal $h$  minimizes this expression.

One possible \emph{heuristic} is to make a mean field approximation and instead minimize $\langle (\hat{Z}_i - h)^2\rangle  \langle (\hat{Z}_{i+1} - h)^2 \rangle$ by setting $h = \langle \hat{Z}_i \rangle$. 
With this choice the onsite term of Eq.~\eqref{eq:mpo_shift} is $\hat{D} = h^2 - 2 h \hat{Z}$, the mean field Hamiltonian.
Since $W^\II$ treats $\hat{D}$ exactly, it's not surprising this can reduce the error. 

To generalize this heuristic mean field criteria, we can always choose the MPO for $H$ such that the decomposition of Eq.~\ref{eq:splitH} satisfies $\braket{h_{R_i, a_i}} = \braket{h_{L_i, a_i}} = 0$ across each bond by shuffling the mean field component into $H_{L_i}, H_{R_i}$.
Then the errors in $\hat{W}^\II$ at 2\textsuperscript{nd}-order will depend only on the connected part of $\sum_{a_i} h_{L_i, a_i} h_{R_i, a_i}$.
For many relevant models, such as a Heisenberg spin model, this heuristic does not help since $\braket{h_{R_i, a_i}} = 0$ due to the SU(2) symmetry of $\mathbf{S}$.
But for a model with a long-ranged density-density interaction like $\tfrac{1}{2} \sum_{x, y} n_x V(x - y) n_y$, the mean field approach will treat the `direct' part of the evolution, $\sum_{x, y} n_x V(x - y) \langle n_y \rangle$, exactly.

\section{Analytical expressions for dynamical correlation functions of Haldane-Shastry spin chain}
\label{app:HS}

We provide here the expression found by Haldane and Zirnbauer~\cite{HaldaneZirnbauer1993} for the ground-state dynamical correlations
\begin{align}
	G^{ab}_{mn}(t,t^\prime) \equiv \langle 0 | S^a_m(t) S^b_n(t^\prime) | 0 \rangle
\end{align}
of the Haldane-Shastry spin chain \cite{Haldane88,Shastry88} with Hamiltonian
\begin{align}
	H_\textrm{HS} = J \sum_{m < n, a} \frac{S^a_m S^a_n}{|m-n|^2} .
\end{align}
(The superscript of $S$ operators denote the spin direction and the subscript denote the lattice site.)
The arguments leading to the forms below are somewhat involved and we refer the reader to the original paper for details.  In the following $\hbar=1$.  $G^{ab}_{mn}$ is diagonal in spin indices, and translation invariance allows us to define
\begin{align}
G^{ab}_{mn}(t,t^\prime)  = \frac{1}{4} \delta^{ab} (-1)^{m-n} C(m-n,t-t^\prime).
\end{align}
The function $C(x,t)$ is related to the spinon spectrum in the solution for the ground-state wavefunction and can be simplified to two integrals:
\begin{align}
	C(x,t) = \frac{1}{4} \int_{-1}^1\!\!d\lambda_1 \int_{-1}^1\!\!d\lambda_2\,
		e^{i \pi \lambda_1 \lambda_2 x - \frac{\pi v t}{2} ({\lambda_1}^2 + {\lambda_2}^2 - 2 {\lambda_1}^2 {\lambda_2}^2)}.
\end{align}
Here $v$ is the spinon velocity, $v = \pi J / 2$, and the prefactor of 1/4 can be understood by noting that $C(0,0) = 1$ as $(S^a)^2 = 1/4$ for each spin direction $a$.
The numerical integrations used to obtain the comparison curves in Fig.~\ref{fig:czz} are straightforward and were carried out using commerical software.

\clearpage

\fi

\end{document}